\documentstyle[12pt]{article}

\newenvironment{indention}[1]{\par%
\addtolength{\leftskip}{#1}%
\begingroup}{\endgroup \par}

\def\ev{{\hbox{eV}}}
\def\kev{{\hbox{keV}}}
\def\mev{{\hbox{MeV}}}
\def\gev{{\hbox{GeV}}}
\def\tev{{\hbox{TeV}}}
\def\lsim{\mathrel{\lower2.5pt\vbox{\lineskip=0pt\baselineskip=0pt
           \hbox{$<$}\hbox{$\sim$}}}}
\def\gsim{\mathrel{\lower2.5pt\vbox{\lineskip=0pt\baselineskip=0pt
           \hbox{$>$}\hbox{$\sim$}}}}

\begin{document}
\baselineskip 7.2mm

\begin{titlepage}

\begin{flushright}
DPNU-94-02\\
AUE-01-94 \\
January \ 1994 \\
\end{flushright}

\vspace {1cm}

\begin{center}
{\Large{\bf How Can We Obtain A Large Majorana-\\
                          Mass in Calabi-Yau Models ?}}

\vspace {1cm}

Naoyuki HABA,${}^1$ \  Chuichiro HATTORI,${}^2$ \
                                      Masahisa MATSUDA,${}^3$ \\
Takeo MATSUOKA${}^1$ and Daizo MOCHINAGA${}^1$ \\
\vspace {3mm}
${}^1${\it Department of Physics, Nagoya University \\
Nagoya, JAPAN 464-01} \\

${}^2${\it Department of Physics, Aichi Institute of Technology \\
Toyota, Aichi, JAPAN 470-03} \\

${}^3${\it Department of Physics and Astronomy \\
Aichi University of Education \\
Kariya, Aichi, JAPAN 448}

\end{center}

\vspace {1cm}


\begin{abstract}
In a certain type of Calabi-Yau superstring models
it is clarified that the symmetry breaking occurs
by stages at two large intermediate energy scales
and that two large intermediate scales induce
large Majorana-masses of right-handed neutrinos.
Peculiar structure of the effective nonrenormalizable
interactions is crucial in the models.
In this scheme Majorana-masses possibly
amount to $O(10^{9 \sim 10}\gev)$ and
see-saw mechanism is at work for neutrinos.
Based on this scheme we propose a viable model
which explains the smallness of masses for three
kind of neutrinos $\nu _e, \nu _{\mu} \
{\rm and}\  \nu _{\tau}$.
Special forms of the nonrenormalizable interactions
can be understood as a consequence of an appropriate
discrete symmetry of the compactified manifold.

\end{abstract}

\end{titlepage}


\section{Introduction}

While superstring theory is the only known candidate of consistent
unification of all fundamental interactions,
untill now we have not succeeded in selecting
a true string vacuum theoretically.
This is because we are lacking a means of addressing
the non-perturbative problems.
In such situation of superstring theory
it is valuable to clarify
how to connect superstring theory with the standard model
and to understand phenomenological implications
of the effective theory from superstring theory.
As a matter of fact, by using phenomenological requirements on
superstring-derived models we can classify
the string vacua corresponding to a huge number of distinct
classical solutions.
It is expected that further study along this point of view
provides an important clue to find a true string vacuum.

In Calabi-Yau superstring models,
unlike the standard gauge group $G_{st} = SU(3)_c \\
\times SU(2)_L \times U(1)_Y$ with rank-four,
the gauge group is rank-six or rank-five at
the compactification scale $M_C$\cite{gaugesym}.
In the followings we discuss rank-six models
coming from abelian flux breaking.
Consequently, there should exist two intermediate
energy scales of symmetry breaking between
the compactification scale and the electroweak scale.
In Calabi-Yau models there appear extra
matter fields which are not contained in
the minimal supersymmetric standard model.
We generally have $G_{st}$-neutral but $E_6$-charged chiral
superfields and their mirror chiral superfields.
Concretely we get $SO(10)$-singlet chiral superfields
and $SU(5)$-singlet chiral superfields
(right-handed neutrino $\nu_R^c$)
denoted as $S$ and $N$, respectively,
which belong to {\bf 27}-representation of $E_6$.
Some of these $G_{st}$-neutral
matter fields have to develop non-vanishing
vacuum expectation values(VEVs)
$\langle S \rangle $ and $\langle N \rangle $
at the intermediate energy scales
in order to connect the Calabi-Yau models with
the standard model.

To solve the so-called hierarchy problem,
it is natural that the supersymmetry(susy) is preserved
down to an energy scale as low as $O(10^3\gev)$.
{}From phenomenological point of view it is well known
that there are at least two large energy scales
between the Planck scale and the soft susy
breaking scale $m_{SUSY}=O(10^3\gev)$.
These scales are concerned with the proton decay and
a large Majorana-mass(M-mass) of the right-handed neutrino.

As for the former subject, in Calabi-Yau models
the lifetime of proton is determined by
the magnitude of $ \langle S \rangle $,
because the superfield $S$ participates in
a Yukawa interaction with leptoquark chiral superfields.
To be consistent with the proton stability,
it is normally required that
$\langle S \rangle  \geq O(10^{16}\gev)$.
Although this condition can be somewhat relieved
provided that the sparticle spectrum is tuned adequately,
even in the case $\langle S \rangle  \geq O(10^{14}\gev)$
is required
\cite{Nath}.

The latter subject is related to see-saw mechanism.
Experimentally neutrino masses are so small compared with
quark masses and charged lepton masses
\cite{seesaw}.
See-saw mechanism provides an interesting solution for the neutrino
mass problem by introducing large M-masses for
right-handed neutrinos.
If we take the solar neutrino problem seriously, the M-mass
of the right-handed neutrino should be of order $10^{9 \sim 12}$\gev
\cite{neutrino}
\cite{SMIRNOV}.
Also this large M-mass is compatible with the cosmological bound
for stable light neutrinos
\cite{Cowsik}.
Since a non-vanishing $\langle N \rangle $ implies
the lepton number violation,
the magnitude of $\langle N \rangle $ seems to be
closely linked to a M-mass of the right-handed neutrino.
A large M-mass suggests a large value of
$\langle N \rangle$.

When $\langle S \rangle, \langle N \rangle
\gg m_{SUSY}$,
we have to make the $D$-terms vanish at such large scales
$\langle S \rangle$ and $\langle N \rangle$.
This is realized by setting
$\langle S \rangle = \langle \overline S \rangle $ and
$ \langle N \rangle = \langle \overline N \rangle $,
where $\overline S$ and $\overline N$ stand for
mirror chiral superfields of $S$ and $N$, respectively.
How can we derive such large intermediate scales in Calabi-Yau
superstring models?
The discrete symmetry of the compactified manifold possibly
accomplishes this desired situation
\cite{discrete}.
In superstring models there exist effective
non-renormalizable(NR) terms in the superpotential.
The order of magnitudes of $\langle S \rangle $ and
$\langle N \rangle $ are governed by these NR terms.
Along this fascinating line
the problems of two large intermediate scales of
symmetry breaking have been studied
first by Masip
\cite{Masip}.
In the analysis general structure of the scalarpotential
has not been sufficiently clarified.
So conditions on the NR terms for the presence of
two large intermediate scales and of
a large M-mass should be studied.

In this paper, we study the NR terms in the superpotential
which satisfy the following two requirements.
The first one is the presence of two large
intermediate energy scales of symmetry breaking.
The second one is the presence of a large M-mass
of $O(10^{9 \sim 12}\gev)$.
The solutions which meet these requirements are found
only in the case when the NR terms are of special forms.
Concretely, the NR interactions of
$S, N$ and $\overline S, \overline N$
are of the form
\begin{equation}
    W_{NR} = M_C{}^3 \lambda _1\,\biggl[
                 \left( \frac {S\overline S}{M_C{}^2} \right)^{2k}
             + k  \left( \frac {N\overline N}
                                   {b^2\,M_C{}^2} \right)^2
             - 2c \left( \frac {S\overline S}{M_C{}^2} \right)^k
                    \left( \frac {N\overline N}
                                   {b^2\,M_C{}^2} \right)
                                      \biggr]
\end{equation}
with $k = 3,4,\ldots $ and $0 < c < \sqrt{2k}$
and $c \neq \sqrt{k}$,
where $\lambda _1$ and $b$ are real constants of $O(1)$.
As a result we have two large intermediate scales
\begin{equation}
        \langle S \rangle  \geq O(10^{16}\gev),
         \ \ \ \ \ O(10^{15}\gev) \geq
                         \langle N \rangle  \geq O(10^{13}\gev)
\end{equation}
and a M-mass of right-handed neutrino becomes
\begin{equation}
        M_M \sim m_{SUSY}
                 \left( \frac {\langle S \rangle }
                              {\langle N \rangle }\right)^2.
\end{equation}
Its numerical value possibly
amounts to $O(10^{9 \sim 10}\gev)$.
Thus see-saw mechanism is at work and
this large M-mass solves the solar neutrino problem.
The main results have been presented in the previous paper
by the present authors
\cite{Ours}.

This paper is organized as follows.
In Sec. 2 we discuss the connection between the NR
interactions and intermediate scales of symmetry breaking.
In the presence of the NR interactions we get
a M-mass matrix by means of
minimization conditions of the scalarpotential.
We require solutions to imply the existence of
two large intermediate scales and of a large M-mass.
In Sec. 3 we look for solutions which meet the requirements.
As a consequence, special types of the NR terms are selected.
The solutions obtained there correspond to a local minimum of
the scalarpotential but not necessarily to the absolute minimum.
The structure of the scalarpotential is studied in detail
for the special types of the NR interactions in Sec. 4.
Under an adequate condition it is shown that the desirable
solution represents the absolute minimum of the scalarpotential.
M-masses are obtained concretely.
To get a M-mass with $O(10^{9\sim 10}\gev)$,
the form of the NR terms are further sorted.
Taking the generation degree of freedom into account,
in Sec. 5 we propose a viable model which explains
the smallness of masses for three kind of
neutrinos $\nu _e, \nu _{\mu} \ {\rm and}\ \nu _{\tau}$.
The final section is devoted to summary and discussion.


\section{Intermediate Scales of Symmetry Breaking}

Before examining in the scheme that
$S, \overline S$ and $N, \overline N$ appear in the
massless spectra at the compactification scale $M_C$,
for illustration we first study the NR
interactions coming from only a pair of $S$ and
$\overline S$ chiral superfields.
The NR terms in the superpotential are of the form
\begin{equation}
        W_{NR} = \sum_{p=2}^{\infty} \lambda_p
                                 M_C{}^{3-2p}(S\overline S)^p,
\end{equation}
where dimensionless coupling $\lambda _p$'s are of order one.
However, if the compactified manifold has a specific type of
discrete symmetry, some of $\lambda _p$'s become vanishing.
When we denote the lowest number of $p$ as $n$, the NR terms
are approximately written as
\begin{equation}
     W_{NR} \cong \lambda_n M_C{}^{3-2n}(S\overline S)^n,
\end{equation}
because the terms with larger $p$ are suppressed by
the inverse power of $M_C$ at low energies.
In the three-generation model obtained from
the Tian-Yau manifold or the Schimmrigk manifold
we have $n = 2, \ 3$
\cite{threege1} \cite{threege2} \cite{threege3}.
While in the four-generation model with the high
discrete symmetry $S_5 \times Z_5{}^5$,
this symmetry leads to $n = 4$
\cite{discrete}.

To maintain susy down to a TeV scale,
the scalarpotential should satisfy $F$-flatness and $D$-flatness
conditions at the large intermediate scale.
Then we have to set
$\langle S \rangle =\langle \overline S \rangle $.
As far as $D$-terms are concerned, the VEV can be taken
as large as we want.
Incorporating the soft susy breaking terms,
we have the scalarpotential
\begin{eqnarray}
       V &=& n^2 \lambda _n{ }^2 M_C{ }^{6-4n}
             \left( \vert S\vert ^{2(n-1)}
                                 \vert \overline S\vert ^{2n}
             +\vert S\vert ^{2n}
                         \vert \overline S\vert ^{2(n-1)} \right)
                                  \nonumber \\
         & & \mbox{ } + {\frac {1}{2}}
                            \sum _{\alpha }g_{\alpha }{}^2
              \left( S^{\dag }T_{\alpha }S
                  - \overline S^{\dag }
                                 T_{\alpha }\overline S \right)^2
                       + V_{soft}, \\
       V_{soft} &=&  m_S{}^2\,\vert S\vert ^2
                       + m_{\overline S}{}^2\,
                                    \vert {\overline S}\vert^2,
\end{eqnarray}
where the $T_{\alpha }$ are Lie algebra generators and
$m_S{}^2$ and $m_{\overline S}{}^2$ are
the running scalar masses squared
associated with the soft susy breaking.
$S$ and $\overline S$ develop nonzero VEVs
when $m_S{}^2+m_{\overline S}{}^2  < 0$.
In the renormalization group analysis
it has been proven that $m_S{}^2+m_{\overline S}{}^2$
possibly becomes negative at the large
intermediate scale $O(10^{16}\gev)$
\cite{Zoglin}.
By minimizing $V$, we obtain the VEVs as
\begin{equation}
       \langle S \rangle  \simeq \langle \overline S \rangle
                  \sim  M_C\left(
                  \frac {\sqrt {-m_S^2}}{M_C}\right)^{1/2(n-1)}.
\end{equation}
The difference $\langle S \rangle - \langle \overline S \rangle$
is negligibly small and
we put $m_{S}{}^2 = m_{\overline S}{}^2$ approximately.
In the case $n=3, \ 4$ the intermediate energy scale becomes
$\langle S \rangle  \simeq \langle \overline S \rangle
\sim O(10^{14}\gev)$, $O(10^{16}\gev)$, respectively,
for $M_C = 10^{18\sim 19}$\gev.
If $n = 2$, then we have $\langle S \rangle  \sim 10^{11}$\gev,
which leads to the fast proton decay.
Through the super-Higgs mechanism,
the $(S - \overline S) /\sqrt {2}$ are absorbed into a massive
vector superfield with its mass of
$O(g_{\alpha } \langle S \rangle )$.
The component $(S + \overline S) /\sqrt {2}$ have masses
of order $O(10^3\gev)$ irrespectively of $n$.
In addition to $\langle S \rangle $ and
$\langle \overline S \rangle $,
we need $\langle N \rangle $ and
$\langle \overline N \rangle $
in order to get sufficiently large M-masses
relative to the soft susy breaking scale.

Next we turn to investigate the case in which
the NR terms consist of pairs of $S$, $N$ and
$\overline S$, $\overline N$ chiral superfields,
provided that there appear $S$, $N$ and
$\overline S$, $\overline N$ superfields
in suitable Calabi-Yau models.
Here we assume the NR interactions
\begin{equation}
    W_{NR} = M_C{}^3 \,\biggl[\lambda _1\,
                           \frac {(S\overline S)^n}{M_C{}^{2n}}
                    + \lambda _2\,
                           \frac {(N\overline N)^m}{M_C{}^{2m}}
                    + \lambda _3\,
                        \frac {(S\overline S)^k(N\overline N)^l}
                                      {M_C{}^{2(k+l)}}\biggr],
\end{equation}
where $n,m,k$ and $l$ are integers with
\begin{equation}
          n > k \geq 1, \ \ \ \ \ m > l \geq 1
\end{equation}
and $\lambda_i$'s are real constants of $O(1)$.
In certain types of Calabi-Yau models it is plausible that
the exponents $n, m, k$ and $l$ are settled
on appropriate values due to the discrete symmetry of
the Calabi-Yau manifold.
In this scheme we potentially derive two intermediate
energy scales of symmetry breaking
and possibly have a large M-mass.
By minimizing the scalarpotential including
the soft susy breaking terms
\begin{equation}
       V_{soft} =   m_S{}^2\,\vert S\vert ^2
                      + m_{\overline S}{}^2\,
                                 \vert {\overline S}\vert^2
                   + m_N{}^2\,\vert N\vert ^2
                      + m_{\overline N}{}^2\,
                                 \vert {\overline N}\vert^2,
\end{equation}
we can determine the energy scales of symmetry breaking,
that is, $\langle S \rangle$ and $\langle N \rangle$.
The scalar mass parameters $m_S{}^2$ and $m_N{}^2$
evolve according to the renormalization group equations.
As shown in ref.\cite{Zoglin},
we expect that
$m_S{}^2$ becomes negative at the large intermediate scale($M_I$).
On the other hand, it is natural to expect that
$m_N{}^2$ remains still positive
at $M_I$ scale,
because quantum numbers and Yukawa interactions of
$N$ and $\overline N$ are quite different
from those of $S$ and $\overline S$.
For this reason we consider the case
$|\langle S \rangle | \gg |\langle N \rangle |$.
Hereafter we take $m_S{}^2(m_{\overline S}{}^2) < 0$
and $m_N{}^2(m_{\overline N}{}^2) > 0$
at $M_I$ scale.
However, the sign of $m_N{}^2$ is not crucial
in the following discussions.
{}From the $D$-flatness condition
we get $|\langle S \rangle| =
|\langle \overline S \rangle|$
and $|\langle N \rangle| = |\langle \overline N \rangle|$
in the approximation $m_S{}^2 = m_{\overline S}{}^2$
and $m_N{}^2 = m_{\overline N}{}^2$.
Here we assume that the VEVs are expressed as
\begin{equation}
       \langle S \rangle = \langle \overline S \rangle
                                     = M_C\, x, \ \ \ \ \
       \langle N \rangle = \langle \overline N \rangle = M_C\, y.
\end{equation}
Without loss of generality $x$ is taken as real for simplicity.
For convenience' sake, instead of $\lambda _i$'s
we use the parameters $a,b$ and $c$ defined as
\begin{equation}
       \lambda _1 = \frac {a}{n}, \ \ \ \ \
       \lambda _2 = \frac {a}{m} \,b^{-2m}, \ \ \ \ \
       \lambda _3 = -\frac {ac}{kl} \,b^{-2l},
\end{equation}
where $a$ is real.
When $\lambda _2/\lambda _1 > 0$,
$b$ and $c$ can be put as real.
For positive $c$ there possibly exist solutions
with real $y$ as seen later.
In the case with negative $c$, if and only if $m$ and $l$
are even and odd, respectively,
we can reduce this case to the case with positive $c$
by redefining the fields $N$ and $\overline N$
attached by a phase factor $i$ as
$N$ and $\overline N$.
When $\lambda _2/\lambda _1 < 0$,
$b$ becomes complex.
However, if we redefine the fields
$N$ and $\overline N$ multiplied by an adequate
phase factor as $N$ and $\overline N$
and then if $c$ becomes real,
this case can be again reduced to the above-mentioned case.
Otherwise, we do not have desirable solutions.
In what follows we put $b$ and $c$ as real and positive
and then $y$ is taken as real.
Let us introduce dimensionless real functions $f$ and $g$ :
\begin{eqnarray}
      f(x,y) & \equiv &
          M_C{}^{-2} \left. \frac {\partial W}{\partial S}\right|
                             = X + Z_f, \nonumber \\
      g(x,y) & \equiv &
          M_C{}^{-2} \left. \frac {\partial W}{\partial N}\right|
                             = Y + Z_g
\end{eqnarray}
with
\begin{eqnarray}
    & &   X   =  a x^{2n-1}, \ \ \ \ \ \ \ \ \ \ \
             Z_f = - \frac {ac}{l} x^{2k-1}
                             \left( \frac {y}{b}\right)^{2l},
                                                 \nonumber \\
    & &   Y   =  \frac {a}{b} \left(\frac
                                {y}{b} \right)^{2m-1}, \ \ \ \ \
             Z_g = - \frac {ac}{kb} x^{2k}
                             \left( \frac {y}{b} \right)^{2l-1},
\label{eqn:xyz}
\end{eqnarray}
where $\ldots \vert $ means the values at
$S = \overline S = \langle S \rangle$
and $N = \overline N = \langle N \rangle$.
By using the $D$-flatness condition we have the scalarpotential
\begin{equation}
       \frac {1}{2} M_C{}^{-4}V\vert = f(x,y)^2 + g(x,y)^2
                            - \rho _x{}^2x^2 + \rho _y{}^2y^2
\label{eqn:v}
\end{equation}
with
\begin{equation}
        \rho _x{}^2 = - \frac {m_S{}^2}{M_C{}^2}\ (>0), \ \ \ \ \
        \rho _y{}^2 =   \frac {m_N{}^2}{M_C{}^2}\ (>0).
\end{equation}
Since $\rho _x$ and $\rho _y$ are of order $O(10^{-(15 \sim 16)})$,
hereafter we often denote $\rho _x$ and $\rho _y$ simply as
a positive parameter $\rho (= O(m_{SUSY}/M_C))$ together.

We are going to carry out the minimization
of the scalarpotential $V$.
Since the scalarpotential is symmetric under the reflection
$x \rightarrow -x$ and/or $y \rightarrow -y$,
it is sufficient for us to consider
only the first quadrant in the $x$-$y$ plane.
The solution of interest here is the one which implies
two large intermediate scales of symmetry breaking
with $\langle S\rangle \gg \langle N\rangle \gg m_{SUSY}$.
At the absolute minimum stationary conditions
\begin{equation}
      \left.\frac {\partial V}{\partial S} \right| =
       \left.\frac {\partial V}
                     {\partial \overline S} \right| =
        \left.\frac {\partial V}{\partial N} \right| =
         \left.\frac {\partial V}
                     {\partial \overline N} \right| = 0
\end{equation}
have to be satisfied.
These conditions are expressed as
\begin {eqnarray}
        f\,f_x + g\,g_x - \rho _x{}^2\,x = 0, \nonumber \\
        f\,f_y + g\,g_y + \rho _y{}^2\,y = 0,
\label{eqn:station}
\end{eqnarray}
where $f_x = {\partial f}/{\partial x}$ and so forth.
More explicitly, we have
\begin{eqnarray}
     & & f_x = \frac {1}{x} [(2n-1)X + (2k-1)Z_f], \nonumber \\
     & & f_y = g_x = \frac {2l}{y} Z_f = \frac {2k}{x} Z_g, \\
     & & g_y = \frac {1}{y} [(2m-1)Y + (2l-1)Z_g]. \nonumber
\end{eqnarray}

For $S, \overline S$ and $N, \overline N$
the mass matrix is given by
\begin{eqnarray}
     \left(
     \begin{array}{cccc}
         W_{SS}|  &  W_{S\overline S}|  &
                      W_{SN}|  &  W_{S\overline N}| \\
         W_{\overline SS}|  &  W_{\overline S\overline S}|  &
               W_{\overline SN}|  &  W_{\overline S\overline N}| \\
         W_{NS}|  &  W_{N\overline S}|  &
               W_{NN}|  &  W_{N\overline N}| \\
         W_{\overline NS}|  &  W_{\overline N\overline S}|  &
               W_{\overline NN}|  &  W_{\overline N\overline N}| \\
     \end{array}
     \right),
\end{eqnarray}
where $W_{SS} = \partial ^2W / \partial S^2$ and so forth.
Through the super-Higgs mechanism
the components $(S-\overline S)/\sqrt 2$ and
$(N-\overline N)/\sqrt 2$ are absorbed by vector superfields
which then become massive with masses
$g_{\alpha }\langle S \rangle $ and
$g_{\alpha }\langle N \rangle $, respectively.
The remaining components $(S+\overline S)/\sqrt 2$ and
$(N+\overline N)/\sqrt 2$ become Majorana superfields.
The mass matrix for $(S+\overline S)/\sqrt 2$ and
$(N+\overline N)/\sqrt 2$ denoted as $M_C \, A$ is of the form
\begin{eqnarray}
     M_C \, A & \equiv &
       \left(
       \begin{array}{cc}
           W_{SS}| + W_{S\overline S}|  &  2W_{SN}| \\
           2W_{SN}|  &  W_{NN}| + W_{N\overline N}|
       \end{array}
       \right) \nonumber \\
     & = & M_C
       \left(
       \begin{array}{cc}
           f_x  &  g_x  \\
           f_y  &  g_y
       \end{array}
       \right)
\label{eqn:majom}
\end{eqnarray}
with $g_x = f_y$.
Here we used the relations
\begin{eqnarray}
       & &  W_{SS}| = W_{\overline S \overline S}|, \ \ \ \ \
              W_{NN}| = W_{\overline N \overline N}|,
                                                    \nonumber \\
       & &  W_{SN}| = W_{S \overline N}| =
              W_{\overline SN}| = W_{\overline S \overline N}|.
\end{eqnarray}
Since the matrix $A$ is real and symmetric,
we can diagonalize this matrix via an orthogonal transformation.
By using the matrix $A$, we can rewrite the stationary conditions
Eq. (\ref{eqn:station}) in the matrix form
\begin{eqnarray}
      A\, \left(
          \begin{array}{c}
               f \\
               g
          \end{array}
          \right)
        = \left(
          \begin{array}{c}
                \rho _x{}^2 x \\
              - \rho _y{}^2 y
          \end{array}
          \right).
\label{eqn:stcon}
\end{eqnarray}

In the next section we solve
Eq. (\ref{eqn:stcon}) in order to find the absolute minimum.
We look for the solution in which $x \gg y \neq 0$
and also a M-mass becomes
sufficiently large relative to $O(10^3\gev)$.
We obtain the constraint on the NR terms
for the existence of desirable solutions.
The constraint yields a relation among the exponents
$n, m, k$ and $l$.


\section{Solutions with A Large Majorana-Mass}

We are now going to find a solution which corresponds to
the absolute minimum with $x \gg y \neq 0$.
Since the value of the scalarpotential
should be negative at the point,
we have
\begin{equation}
        f^2 + g^2  <  \rho _x{}^2x^2 - \rho _y{}^2y^2
                                            \sim \rho ^2\,x^2
\label{eqn:rhoxx}
\end{equation}
for the solution,
where we used the relation $x^2 \gg y^2$.
Furthermore, it can be shown that
\begin{equation}
        f^2 + g^2 = O(\rho ^2x^2).
\label{eqn:rhox}
\end{equation}
If it were not for the case, we have
$ f^2 + g^2 \ll \rho ^2\,x^2 $.
This implies that
$ |f|, \ |g| \ll \rho \,x $.
On the other hand, from Eq. (\ref{eqn:station}) we get
$ f\,xf_x + g\,xg_x \sim \rho ^2 x^2 $.
Then it is impossible that both $|f\,xf_x|$
and $|g\,xg_x|$ are smaller than $O(\rho ^2x^2)$.

If $|f\,xf_x| \gsim \rho ^2x^2$,
we have $|xf_x| \gg \rho x \gg |f|$.
This means that the cancellation of the leading terms
of $X$ and $Z_f$ occurs in $f$.
In this case we get
$ |Z_f| \sim |xf_x| \gg \rho \,x $.
Thus
$ |Z_g| \sim (x/y)|Z_f| \gg \rho x^2/y \gg \rho x \gg |g| $.
This means that the cancellation occurs
also between $Y$ and $Z_g$ in $g$.
However, the cancellation of the leading terms both in $f$
and $g$ results in a high degree of fine tuning
which we consider unlikely.
In fact, by eliminating $x$ in the relations
$f = X + Z_f \sim 0$ and $g = Y + Z_g \sim 0$
we obtain
\begin{equation}
     \left(\frac {y}{b}\right)^{2(nm-nl-mk)} =
                        \left(\frac{k}{l}\right)^k
                          \left(\frac {k}{c}\right)^n
\end{equation}
at the leading order.
In the case the exponent $mn-nl-mk \neq 0$,
$x$ and $y$ turn out to be
expressed as functions only of $b$ and $c$.
By substituting these into Eq. (\ref{eqn:station})
we have relations between $b, c$ and $\rho _x, \rho _y$.
Parameters $\rho _x$ and $\rho _y$ are the running ones
of the soft susy breaking
determined by the renormalization group equations.
While $b$ and $c$ are coupling constants of
the NR terms in superpotential.
Therefore, these relations imply a fine tuning
which we consider unlikely.
In the case $mn-nl-mk=0$, $c$ is fixed to a specific value.
However, it is also unlikely that such a special value of $c$
is derived from the discrete symmetry of the compactified
manifold.

Next we consider the case $|f\,xf_x| \ll \rho ^2x^2$ and then
$|g\,xg_x| \sim \rho ^2x^2$.
Similarly to the above argument, we get
$ |xg_x| \gg \rho x \gg |g| $.
Then the cancellation of the leading terms of $Y$ and $Z_g$
have to take place in $g$.
Since this means $|yg_y| \sim |xg_x|$,
we obtain $|g\,yg_y| \sim |g\,xg_x| \sim \rho ^2x^2$.
While, from Eq. (\ref{eqn:station}) we have
$ f\,yf_y + g\,yg_y \sim \rho ^2 y^2 $.
In order to satisfy this relation under $x^2 \gg y^2$,
$|f\,yf_y| \sim |g\,yg_y| \sim \rho ^2x^2$ and
the leading terms of $f\,yf_y$ and $g\,yg_y$ have
to cancel out with each other.
In this case we get $|yf_y| \gg \rho x \gg |f|$ and
then $|Z_f| \sim |xf_x| \sim |yf_y|$.
Thus $|f\,xf_x| \sim \rho ^2x^2$.
This  contradicts with the relation supposed here.
Therefore, we obtain the relation (\ref{eqn:rhox}).

Next we show that $|f| = O(\rho x)$ and that
only one M-mass possibly becomes large
compared with $m_{SUSY}$.
Through an orthogonal transformation we carry out the
diagonalization of the matrix $A$ as
\begin{equation}
      U\, A\, U^{-1} =
  \left(
  \begin{array}{cc}
          \omega _1    &      0      \\
              0        &  \omega _2
  \end{array}
  \right),
\label{eqn:ad}
\end{equation}
where
\begin{equation}
      U =
  \left(
  \begin{array}{cc}
          \cos \theta    &   -\sin \theta    \\
          \sin \theta    &    \cos \theta
  \end{array}
  \right).
\end{equation}
Then M-masses are $M_C|\omega _1|$ and $M_C|\omega _2|$.
We require that at least one of $|\omega _1|$
and $|\omega _2|$ is sufficiently larger
than $O(\rho )=O(m_{SUSY}/M_C)$.
However, it is impossible that both $|\omega _1|$
and $|\omega _2|$ are larger than $O(\rho )$.
To see this, let us suppose for a moment that both $|\omega _1|$
and $|\omega _2|$ are larger than $O(\rho )$,
i.e., $O(\rho ) \ll |\omega _1| \leq |\omega _2|$.
{}From Eq. (\ref{eqn:stcon}) we have
\begin{eqnarray}
     \left( \rho _x{}^2\,x\right)^2
                 + \left( \rho _y{}^2\,y\right)^2
     & = &
       \left(
       \begin{array}{cc}
              f   &  g
       \end{array}
       \right)
              A^T\,A
       \left(
       \begin{array}{c}
              f  \\
              g
       \end{array}
       \right)                                      \nonumber \\
     & = & \omega _1{}^2\,\left(f\cos \theta
                             - g\sin \theta \right)^2
           + \omega _2{}^2\,\left(f\sin \theta
                             + g\cos \theta \right)^2
                                                    \nonumber \\
     & \geq & \omega _1{}^2\left( f^2 + g^2 \right). \nonumber \\
     & \sim & \omega _1{}^2\rho ^2x^2,  \nonumber
\end{eqnarray}
where we used Eq. (\ref{eqn:rhox}).
This is inconsistent with the relation supposed here.
Thus we have to be
\begin{equation}
     |\omega _1| \leq O(\rho ), \ \ \ \ \
     |\omega _2| \gg  O(\rho ).
\label{eqn:omega}
\end{equation}

{}From Eq. (\ref{eqn:ad}) $A$ is expressed as
\begin{equation}
      A =
       \left(
       \begin{array}{cc}
           f_x  &  g_x  \\
           f_y  &  g_y
       \end{array}
       \right)
       =
       \left(
       \begin{array}{cc}
          \omega _1\cos ^2\theta  + \omega _2\sin ^2\theta
           &   (\omega _2-\omega _1)\sin \theta \cos \theta   \\
          (\omega _2-\omega _1)\sin \theta \cos \theta
           &   \omega _1\sin ^2\theta  + \omega _2\cos ^2\theta
       \end{array}
       \right).
\label{eqn:a}
\end{equation}
Then we obtain
\begin{eqnarray}
    f_x & = & \omega _1 \cos ^2\theta  +
                          \omega _2 \sin ^2\theta ,
                                 \label{eqn:fx}  \\
    f_y & = & (\omega _2 - \omega _1) \sin \theta \cos \theta.
\label{eqn:fy}
\end{eqnarray}
Unless $|\sin \theta | \ll 1$,
it turns out that
$|\omega _2| \sin^2 \theta \gg |\omega _1| \cos^2 \theta $
because of Eq. (\ref{eqn:omega}).
By using $Z_f \sim yf_y$ and Eq. (\ref{eqn:fy}),
we get
$ |x f_x| \sim |\omega _2| x \sin^2 \theta \gg
             |\omega _2 y \sin \theta \cos \theta | \sim |Z_f| $.
This implies that $|X| \gg |Z_f|$ and then
$ |f| \sim |xf_x| \sim |\omega _2| x \sin^2 \theta
                                   \gg  \rho x $.
This contradicts with Eq. (\ref{eqn:rhox}).
Thus we are led to the inequality
\begin{equation}
     |\sin \theta | \ll 1.
\end{equation}
Without loss of generality we can take $|\theta | \ll 1$.
Then the matrix $A$ is approximated as
\begin{equation}
      A =
       \left(
       \begin{array}{cc}
           f_x  &  g_x  \\
           f_y  &  g_y
       \end{array}
       \right)
       \simeq
       \left(
       \begin{array}{cc}
          \omega _1 + \omega _2 \theta ^2
                             &   \omega _2 \theta   \\
          \omega _2 \theta   &   \omega _2
       \end{array}
       \right).
\label{eqn:aa}
\end{equation}
Combining Eq. (\ref{eqn:stcon}) with this expression we obtain
\begin{eqnarray}
       (\omega _1 + \omega _2 \theta ^2)f +
                   \omega _2 \theta g \simeq \rho _x{}^2 x,
                                          \label{eqn:st1} \\
       \omega _2 \theta f + \omega _2 g \simeq -\rho _y{}^2 y .
\label{eqn:st2}
\end{eqnarray}
Subtracting Eq. (\ref{eqn:st2}) multiplied by $\theta $
from Eq. (\ref{eqn:st1}), we find
\begin{equation}
       \omega _1 f \simeq \rho _x{}^2 x.
\end{equation}
In consideration of Eqs. (\ref{eqn:rhox}) and (\ref{eqn:omega}),
this leads us to
\begin{equation}
        |\omega _1| = O(\rho )
\end{equation}
and $|f| \sim \rho x$.
Therefore, Eqs. (\ref{eqn:st1}) and (\ref{eqn:st2})
are translated as
\begin{eqnarray}
       & & |f| \sim \rho x,      \label{eqn:ff} \\
       & & \theta f + g  \sim
                   -\frac {\rho ^2}{\omega _2} y
\label{eqn:fg}
\end{eqnarray}
with  $|\omega _2| \gg O(\rho )$.

{}From Eq. (\ref{eqn:aa}) we get
\begin{eqnarray}
       & & f_x  \sim  \rho + \omega _2 \theta ^2,
                                       \label{eqn:aaa} \\
       & & f_y    =  g_x \sim \omega _2 \theta ,
                                       \label{eqn:aab} \\
       & & g_y  \sim  \omega _2.
                                       \label{eqn:aac}
\end{eqnarray}
To solve these equations together with Eqs. (\ref{eqn:ff})
and (\ref{eqn:fg}), it is convenient for us to
classify into two cases according to whether or not
the cancellation between the leading terms of $Y$ and $Z_g$
occurs in $g$.
First consider the case when there is no cancellation in $g$.
Taking Eqs. (\ref{eqn:aab}) and (\ref{eqn:aac})
into account, we can compare the magnitude of each term
in Eq.(\ref{eqn:fg}).
On the left hand side of Eq. (\ref{eqn:fg}) $|g|$ is
sufficiently larger than $|\theta f|$, because
$ |g| \gsim |Z_g| \sim |xg_x| \sim |\theta \omega _2|x
                   \gg |\theta |\rho x \sim |\theta f| $.
While the right hand side of Eq. (\ref{eqn:fg}) is much
smaller than $|g|$, i.e.,
$ \left| \rho ^2 y / \omega _2 \right| \ll
        \rho y \ll |\omega _2|y \sim |yg_y| \lsim |g| $,
where we used Eq. (\ref{eqn:aac}).
Then Eq. (\ref{eqn:fg}) can not be satisfied in this case.
Therefore, a cancellation of the leading terms in $g = Y +Z_g$
have to take place and
a cancellation does not occur in $g_y$.
Thus
\begin{equation}
        |xg_x| \sim |Z_g| \sim |Y| \sim |yg_y|.
\end{equation}
Using Eqs. (\ref{eqn:aab}) and (\ref{eqn:aac}), we get
$|\omega _2\theta |x \sim |\omega _2|y$.
This means that
\begin{equation}
      |\theta | \sim \frac {y}{x}.
\end{equation}
{}From Eq. (\ref{eqn:a}) we have
$g_y = \omega _2 + O(\omega _2\theta ^2)$.
Since the next-to-leading term is suppressed by
$\theta ^2$ relative to the leading one,
we can express as
\begin{equation}
        |g| \sim \theta ^2 |Y|.
\label{eqn:epsilon}
\end{equation}
The magnitude of each term in Eq. (\ref{eqn:fg}) is
estimated as
\begin{eqnarray}
     & & |\theta f| \sim |\theta |\rho x \sim \rho y,
                               \label{eqn:thetaf} \\
     & & |g| \sim |\theta ^2 Y| \sim |\theta ^2 yg_y|
                            \sim |\theta ^2 \omega _2|y,
                               \label{eqn:gg} \\
     & & \left| -\frac {\rho ^2}{\omega _2}y\right|
                                   \ll \rho y.
\end{eqnarray}
Consequently, in order that Eq. (\ref{eqn:fg}) holds,
the leading terms of $\theta f$ and $g$ have to cancel
out with each other.
Thus  from Eqs. (\ref{eqn:thetaf}) and (\ref{eqn:gg})
we obtain
\begin{equation}
      \theta ^2 \sim \frac {\rho }{|\omega _2|}.
\end{equation}
Returning to Eqs. (\ref{eqn:aaa}) and (\ref{eqn:aab}),
we get
\begin{equation}
      |f| \sim |X| \sim |Z_f|.
\end{equation}

The conclusion of this section is that
a desirable solution exists only in the case when
a cancellation of the leading terms occurs
in $g$ but not in $f$ and $f_x$.
At the same time $|X| \sim |Z_f|$ should be satisfied.
Combining this with the relation
$|Y| \sim |Z_g|$ and Eq. (\ref{eqn:xyz}),
we find
\begin{equation}
      \frac {n}{m} = \frac {k}{m-l} > 1
\label{eqn:nmij}
\end{equation}
and
\begin{eqnarray}
        & &   x \sim \rho ^{1/2(n-1)},  \\
        & &   y \sim x^{n/m} \sim \rho ^{n/2m(n-1)}.
\end{eqnarray}
Finally, a large M-mass becomes
\begin{equation}
      M_C|\omega _2| \sim m_{SUSY}\left(
                             \frac {x}{y} \right)^2.
\end{equation}
%


\section{Minimization of Scalarpotential}

Although in the previous section we find
desirable solutions,
a question arises as to whether or not
the solution found there represents the absolute
minimum of the scalarpotential.
Then in this section we study the structure of
the scalarpotential concretely and find an additional
condition such that a desirable solution becomes
the absolute minimum of the scalarpotential.
Since we consider the case when
the relation (\ref{eqn:nmij}) is satisfied,
we get $|X| \sim |Z_f|$ and $|Y| \sim |Z_g|$
coincidentally in the region $x^n \sim y^m$.
Solving the stationary condition (\ref{eqn:stcon}),
one can find local minima and saddle points
of the scalarpotential.
In this case, it can be proven for the scalarpotential
with $\rho _y{}^2 > 0$ that
there are the following two or three local minima.
The values of the scalarpotential at these points
are calculable.

\begin{indention}{5mm}
    {\bf Point \ A}: \ \ \ \ \ $(x,y)  =  (x_0,y_0)$.
\begin{equation}
          M_C{}^{-4}V  \cong
                - \frac {4(n-1)}{(2n-1)} \,\rho _x{}^2 x_0{}^2,
\label{eqn:VA}
\end{equation}
\begin{indention}{5mm}
where
\begin{eqnarray}
            & x_0 &= \left(
                      \frac {\rho _x}{\sqrt {2n-1} \,a \,\xi }
                    \right) ^{1/2(n-1)} , \nonumber \\
            & y_0 &= b \left( \frac {c}{k} \right) ^{1/2(m-l)}
                       x_0{}^{k/(m-l)} \ \ \ (\ll x_0), \\
            & \xi &= \left| 1-\frac{k}{l} \left( \frac {c}{k}
                                  \right) ^{n/k} \right|. \nonumber
\end{eqnarray}
\end{indention}
    {\bf Point \ B}: \ \ \ \ \ $(x,y)  =  (x_0',0)$.
\begin{equation}
          M_C{}^{-4}V  \cong
                - \frac {4(n-1)}{(2n-1)} \rho _x{}^2 {x_0'}^2,
\label{eqn:VB}
\end{equation}
\begin{indention}{5mm}
where
\begin{equation}
             x_0' = \left(
                      \frac {\rho _x}{\sqrt {2n-1} \,a }
                    \right) ^{1/2(n-1)}.
\end{equation}
\end{indention}

    {\bf Point \ C}: \ \ \ \ \ $(x,y)  =
                 (x_0',y_0')$ \ \ \ $({\rm only \ for\ } l \geq 2
                                 \ {\rm and} \ 1+R \geq 0)$.
\begin{equation}
          M_C{}^{-4}V  \cong
                - \frac {4(n-1)}{(2n-1)} \,\rho _x{}^2 {x_0'}^2,
\label{eqn:VC}
\end{equation}
\begin{indention}{5mm}
where
\begin{eqnarray}
            & y_0' &= b \left(
                         \frac {k^2b^2}{(2l-1)c}
                            \left( 1 + \sqrt {1+R} \right)
                       \right)^{1/2(l-1)}
                       {x_0'}^{(n-k-1)/(l-1)}
                               \ \ \ (\ll y_0), \nonumber \\
            & R &= - \frac {(2n-1)(2l-1)\rho _y{}^2}
                          {k^2 \rho _x{}^2}\ \ \ (< 0).
\end{eqnarray}
\end{indention}
\end{indention}
Point A is a solution which was obtained in the previous
section and also found by Masip\cite{Masip}.
At this point not only two terms in $g(x,y)$ cancel out
with each other in their leading order
but also the leading term of $f\,f_y$
in Eq. (\ref{eqn:station}) cancels out $g\,g_y$.
In the expansion the ratio of the next-to-leading terms to
the leading ones is $O((y_0/x_0)^2)$.
In the case $l \geq 2$ and $1+R \geq 0$,
Point C becomes a local minimum
but not in the other cases.
Although $x$ and $y$ are non-zero at Point C,
Point C is not a desirable solution because
M-masses are $O(m_{SUSY})$.
In addition to local minima, we also have saddle points
which are located at the origin and the following points.

\begin{indention}{5mm}
    {\bf Point \ D}: \ \ \ \ \ $(x,y)  =  (x_1,y_1)$.
\begin{indention}{5mm}
Where
\begin{eqnarray}
            & x_1 & = \eta ^{n(2m-1)/2k\phi }
                \left( \frac {(2l-1)b\rho _x}
                       {2a\sqrt{k(m-l)(2m-1)}}
                \right)^{m/\phi } \ \ \ (\ll x_0),  \nonumber \\
            & y_1 & = b \left( \frac {(2l-1)b\rho _x}
                       {2a\sqrt{k(m-l)(2m-1)}}
                \right)^{m/\phi } \ \ \ (\ll y_0)
\end{eqnarray}
\begin{indention}{5mm}
with
\end{indention}
\begin{eqnarray}
           & & \eta = \frac {k(2m-1)}{(2l-1)c}, \nonumber \\
           & & \phi = 2mn-m-n.
\end{eqnarray}
\end{indention}

    {\bf Point \ E}: \ \ \ \ \ $(x,y)  =
                 (x_0',y_1')$ \ \ \ $({\rm only \ for \ } l \geq 2
                                 \ {\rm  and } \ 1+R \geq 0)$.
\begin{indention}{5mm}
Where
\begin{eqnarray}
            & y_1' &= b \left(
                         \frac {k^2b^2}{(2l-1)c}
                            \left( 1 - \sqrt {1+R} \right)
                       \right)^{1/2(l-1)}
                       {x_0'}^{(n-k-1)/(l-1)}
                               \ \ \ (\ll y_0).
\end{eqnarray}
\end{indention}
\end{indention}
\noindent
If $\rho _y{}^2 < 0$, we have two local minima at Points A
and B for $l=1$ and at Points A and C for $l \geq 2$.

In comparison of Eq. (\ref{eqn:VA})
with Eqs. (\ref{eqn:VB}) and (\ref{eqn:VC}),
Point A becomes the absolute minimum under
the condition $0 < \xi < 1$.
This condition on $\xi $ is translated as
\begin{equation}
      0 < c < k \left( \frac {2l}{k} \right)^{k/n}
             {\rm and} \ \ c \neq k \left( \frac {l}{k}
                          \right)^{k/n}.
\label{eqn:cc}
\end{equation}
It is worth noting that under this condition the Point A
is the absolute minimum independent of the sign of $m_N{}^2$.
For illustration we show  the behavior of the scalarpotential
for the cases $(n,k,m,l) = (6,3,2,1)$ and $(9,3,3,2)$
in Figs. 1 and 2, respectively.
In these Figures the vertical axis is taken as
\begin{equation}
        v = \left( 2M_C{}^4 \rho _x{}^2 x_0{}^2
                                 \right)^{-1} V + 1
\end{equation}
and instead of $x$ and $y$ the horizontal axes are taken as
$\overline x = (x/x_0)^{n/m}$ and $\overline y = y/y_0 $
so that the point $(\overline x, \overline y) = (1, 1)$
becomes the absolute minimum (Point A).
In the case $(n,k,m,l) = (6,3,2,1)$ the condition
(\ref{eqn:cc}) leads to
$0 < c < \sqrt 6$ and $c \neq \sqrt 3$.
Here we put $a = b = c = 1$ in Fig. 1 and
$a=b=1,\ c=2$ in Fig. 2.
As seen in Fig. 1, local minima (Points A and B)
are located at bottoms of very deep canyons.
This comes from the fact that a curveture along
the direction perpendicular to the line $x^n = y^m$
represents a large M-mass squared.
In the case $m = 2$ and $l = 1$ the canyon is most steep.
In the other cases the slope of the canyon becomes
gentle relative to the case $m = 2$ and $l = 1$.
These situations are seen in Figs. 1 and 2.

\vskip 0.5cm

\begin{center}
 \unitlength=0.8cm
 \begin{picture}(2.5,2.5)
  \thicklines
  \put(0,0){\framebox(3,1){\bf Fig. 1}}
 \end{picture}
\end{center}

\vskip 0.5cm

\begin{center}
 \unitlength=0.8cm
 \begin{picture}(2.5,2.5)
  \thicklines
  \put(0,0){\framebox(3,1){\bf Fig. 2}}
 \end{picture}
\end{center}

\vskip 0.5cm

We are now in a position to evaluate the M-mass matrix for
$(S+\overline S)/\sqrt 2$ and $(N+\overline N)/\sqrt 2$
at the absolute minimum (Point A).
The mass matrix is of the form
\begin{equation}
      M_C\,A = M_C \left(
                   \begin{array}{cc}
                      f_x    &     g_x    \\
                      f_y    &     g_y
                   \end{array}
                   \right)
             = m_{SUSY}
         \left(
         \begin{array}{cc}
             O(1)            &     O( x_0/y_0 )    \\
             O( x_0/y_0 )    &     O( (x_0/y_0)^2 )
         \end{array}
         \right).
\end{equation}
More precisely, the matrix elements are
\begin{eqnarray}
       & & f_x = \rho _x \frac {1}{\sqrt {2n-1} \,\xi }
                  \left[ (2n-1)-(2k-1)\frac {k}{l}
                      \left(\frac {c}{k}\right)^{n/k}
                  \right], \nonumber \\
       & & f_y = g_x = \rho _x \left( \frac {x_0}{y_0} \right)
                  \frac {2k}{\sqrt {2n-1} \,\xi }
                  \left(\frac {c}{k}\right)^{n/k}, \nonumber \\
       & & g_y = \rho _x \left( \frac {x_0}{y_0} \right)^2
                  \frac {2(m-l)}{\sqrt {2n-1} \,\xi }
                      \left(\frac {c}{k}\right)^{n/k}.
\end{eqnarray}
Thus we obtain a large M-mass
\begin{equation}
          M_{N'} = M_C\,\omega_2 =
                 \frac {2(m-l)}{\sqrt {2n-1} \,\xi }
                  \left( c/k\right)^{n/k}
                   \sqrt  {-m_S{}^2} \left( {x_0}/{y_0} \right)^2,
\label{eqn:Mm}
\end{equation}
which is associated with the eigenstate
\begin{equation}
       N' =
         \cos \theta \frac {1}{\sqrt 2}( N + \overline N )
           + \sin \theta \frac {1}{\sqrt 2}( S + \overline S )
\end{equation}
with
\begin{equation}
        \theta = - \frac {k}{(m-l)}
                     \left( \frac {y_0}{x_0} \right).
\end{equation}
The eigenstate with mass $M_C|\omega _1| = O(m_{SUSY})$
is given by
\begin{equation}
      S' =
         -\sin \theta \frac {1}{\sqrt 2}( N + \overline N )
           + \cos \theta \frac {1}{\sqrt 2}( S + \overline S ).
\end{equation}
The enhancement factor $(x_0/y_0)^2$ in Eq. (\ref{eqn:Mm})
depends on $n$ and $m$ as
\begin{equation}
         \left( {x_0}/{y_0} \right)^2 \sim
                   \left( {1}/{\rho _x} \right)^{(n-m)/(n-1)m}
\end{equation}
with $\rho _x{}^{-1} = M_C/\sqrt {-m_S{}^2} = 10^{15\sim 16}$.
Since the exponent $(n-m)/(n-1)m$ decreases with increasing $m$,
we take $m=2$ so as to get a sufficiently large M-mass $M_{N'}$.
Then we have $l=1$ and $n=2k$.
This leads to
\begin{equation}
         M_{N'} = x_0\,O \left( \sqrt {M_C \times m_{SUSY}}
                                                   \right).
\end{equation}
Numerically we obtain
\begin{equation}
         \left( {x_0}/{y_0} \right)^2 = 10^{7 \sim 8}
                        \ \ \ \ \ {\rm for} \ \ n \geq 6
\end{equation}
and the M-mass becomes
\begin{equation}
            M_{N'} = O\left( 10^{9 \sim 10} \gev \right)
\end{equation}
by taking $\sqrt {-m_S{}^2} = O(10^3\gev)$.
Consequently, a large M-mass can be induced from
the NR interactions of
$S, N$ and $\overline S, \overline N$
which are of the form
\begin{equation}
    W_{NR} = M_C{}^3 \lambda _1\,\biggl[
                 \left( \frac {S\overline S}{M_C{}^2} \right)^{n}
             + \frac {n}{2}
                    \left( \frac {N\overline N}
                                   {b^2\,M_C{}^2} \right)^2
             - 2c \left( \frac {S\overline S}{M_C{}^2} \right)^{n/2}
                    \left( \frac {N\overline N}
                                   {b^2\,M_C{}^2} \right)
                                      \biggr]
\label{eqn:WNR}
\end{equation}
with $0 < c < \sqrt{n}$ and $c \neq \sqrt{n/2}$.
For comparison we tabulate the orders of
$\langle S \rangle, \langle N \rangle $ and $M_{N'}$
for several cases of the set $(n,k,m,l)$ in Table I.
As seen in this Table, unless $m=2$ and $l=1$,
$M_{N'}$ attains to only at most $O(10^7\gev)$.
The case $m=2$ and $l=1$,
which leads to $n=2k$,
is indispensable for solving
the solar neutrino problem.

\vskip 0.5cm

\begin{center}
 \unitlength=0.8cm
 \begin{picture}(2.5,2.5)
  \thicklines
  \put(0,0){\framebox(3,1){\bf Table I}}
 \end{picture}
\end{center}

\vskip 0.5cm


\section{ Small See-saw Neutrino Masses}

In the previous sections,
we have constructed a consistent model with two large
intermediate energy scales of symmetry breaking.
The higher energy scale is given
by the VEV $\langle S \rangle = \langle {\overline S}
\rangle = O(10^{16\sim 18}\gev)$ which
can prohibit fast proton decay.
The other energy scale is the VEV
$\langle N \rangle=\langle{\overline N}\rangle=
O(10^{13 \sim 15}\gev)$.
These scales induce the large Majorana neutrino
mass $M_{N'}$ with $O(10^{9 \sim 10}\gev)$.

In this section we propose a viable model
which explains the smallness of three kind of
neutrinos $\nu_e, \nu_\mu {\rm and}\ \nu_\tau$.
This problem could be reduced to see-saw mechanism
\cite{seesaw}.
The present experimental limits on neutrino masses are
given as
\cite{PDG}
\begin{equation}
       m_{\nu_e} < 7.3 \ev, \qquad
       m_{\nu_\mu} < 270 \kev, \qquad
       m_{\nu_\tau} < 35 \mev
\end{equation}
by the laboratory experiments.
On the other hand, recent experiments on solar neutrino and
atomospheric neutrino have given more stringent constraints
on neutrino masses and mixing parameters.
{}From solar neutrino experiments
by Homestake, Kamiokande and recent GALLEX
\cite{SMIRNOV}
\cite{HIRATA}
\cite{GALLEX}
the allowed nonadiabatic narrow MSW band
\cite{neutrino} is
\begin{equation}
       \Delta m^2_{12} \simeq (2.7 \sim 13)
                             \times 10^{-6}{\rm eV}^2 \ , \qquad
       \sin^2{2\,\theta_{12}} \simeq 0.004 \sim 0.013
\end{equation}
for the mixing among the first and
the second generations.
Atomospheric neutrino experiments by Kamiokande
and IMB Collaboration\cite{HIRATA}\cite{IMB}
have shown the depletion of the atmospheric muon-neutrino flux.
The allowed neutrino oscillation parameters have been
given in Ref.\cite{HIRATA}.
{}From this it is expected that the heaviest neutrino mass
among three light neutrinos is $O(10^{-1}\ev)$
and that there is a large mixing of
the muon-neutrino with another neutrino.
If we combine the solar neutrino data with
the atomospheric neutrino ones,
the possible mixing solution is given by
\begin{equation}
       \Delta m^2_{23} \simeq (2 \sim 40)
                       \times 10^{-3} {\rm eV}^2 \ , \qquad
       \sin^2{2\,\theta_{23}} \simeq 0.4 \sim 0.7 \ .
\end{equation}
{}From these results the neutrino masses are
\begin{eqnarray}
       m_{\nu_\mu}  & \simeq & (1.6 \sim 3.6)
                             \times 10^{-3}{\rm eV}, \\
       m_{\nu_\tau} & \simeq & (0.4 \sim 2)
                             \times 10^{-1}{\rm eV},
\end{eqnarray}
provided that $m_{\nu_e} \ll m_{\nu_\mu} \ll m_{\nu_\tau}$.

Here we are going to estimate the magnitude of large
M-masses which lead to sufficiently small
neutrino masses by see-saw mechanism.
To do this, we need to know the Dirac-mass matrix
for neutrinos.
We take two possibilities for the structure of neutrino
Dirac-mass matrix.

One possibility is that the leptonic Dirac-mass matrix is
the same as the quark one from the standpoint of
the quark-lepton unification at the Planck scale.
We take masses of up-quark sector
as Dirac-mass terms of neutrinos.
{}From the above constraints on neutrino masses
we obtain right-handed M-masses
\begin{eqnarray}
         M_{M_2} & \simeq & {m_c^2 \over m_{\nu_\mu}}
                     \sim (0.6 \sim 1.4) \times 10^{12}
                      {\rm GeV},  \\
         M_{M_3} & \simeq & {m_t^2 \over m_{\nu_\tau}}
                     \sim(1 \sim 6) \times 10^{14}{\rm GeV}
\end{eqnarray}
by taking $m_c = 1.4$\gev  \ and  $m_t = 150$\gev.
In this case we are compelled to get mass hierarchy also
for right-handed M-mass matrix.
To obtain a reasonably light neutrinos
by using see-saw mechanism,
we need at least two M-masses of
$O(10^{12}\gev)$ and $O(10^{14}\gev)$
as derived above.
In the present model, it is difficult to obtain
a M-mass as large as $O(10^{14}\gev)$.

The other possiblity is that the structure of
Dirac neutrino masses is the same as the one
of charged lepton masses.
In this case right-handed M-masses become
\begin{eqnarray}
        M_{M_2} & \simeq & {m_\mu^2 \over m_{\nu_\mu}}
                   \sim (0.3 \sim 0.7) \times
                                  10^{10}{\rm GeV}, \\
        M_{M_3} & \simeq & {m_\tau^2 \over m_{\nu_\tau}}
                   \sim (1.6 \sim 8.0) \times
                            10^{10}{\rm GeV}.
\end{eqnarray}
Then a single M-mass scale with $O(10^{10}\gev)$
can reproduce the small neutrino masses consistent
with recent solar and atomospheric neutrino experiments.
Unfortunately, at present there is no theoretical basis
that guarantees the equality
$m_{\nu_i}(Dirac \ mass) \simeq m_{e_i}$,
where $m_{e_i}$ means $i$-th charged lepton mass.
We now have no knowledge about Yukawa couplings
$N_iL_jH_u$ at the compactification scale.
So this similarity between neutral and
charged Dirac-mass terms is an important subject
that we should derive from superstring theory
in the future study.

Now we propose a simple model with three generations
along the scenario given in the previous section.
First we consider a case in which all generations
of right-handed sneutrino $N_i$ develop almost
the same VEV in magnitude, i.e.,
\begin{equation}
        \langle N_i\rangle = O(10^{13 \sim 15}{\rm GeV}).
\end{equation}
This scenario is implemented by substituting
$\Sigma_{i=1}^{3} N_i$ for $N$ in Eq. (\ref{eqn:WNR}).
However, the superpotential contains the Yukawa
interaction term like $N_iL_j H_{\rm u}$,
where the indices $i,j$ mean the generation degree of freedom
and we assume only one generation for Higgs sector
below the scale $\langle S \rangle $.
This term generates the large mixing masses for
$L_jH_{\rm u}$ due to the VEV
 $\langle N_i\rangle=O(10^{13 \sim 15}\gev)$.
So these large mixing masses bring about the large
Dirac-masses for neutrino states and then
this model is inconsistent with the small neutrino masses.

Therefore, as an alternative to the above case,
we next consider the case that
the VEVs become
\begin{eqnarray}
       & & \langle N \rangle = \langle{\overline N}\rangle
             = O(10^{13\sim 15}{\rm GeV}), \nonumber  \\
       & & \langle N_1 \rangle = \langle N_2 \rangle =
                     \langle N_3 \rangle = 0.
\label{eqn:nvev}
\end{eqnarray}
To construct a viable model it is assumed
that we have the Yukawa interactions
$N_iL_j H_{\rm u}$ but not $N L_j H_{\rm u}$.
The NR interactions
\begin{equation}
       W_{NR} = W^{(0)}_{NR} + W^{(1)}_{NR}\,.
\label{eqn:NR}
\end{equation}
implements this situation (\ref{eqn:nvev}), where
\begin{equation}
    W^{(0)}_{NR} = M_C{}^3 \lambda _1\,
                  \biggl[
                   \left( \frac {S\overline S}{M_C{}^2}
                                              \right)^n
                   + \frac {n}{2} \left( \frac {N\overline N}
                                   {b^2\,M_C{}^2} \right)^2
                   - 2c \left( \frac {S\overline S}{M_C{}^2}
                                                 \right)^{n/2}
                   \left( \frac {N\overline N}
                                   {b^2\,M_C{}^2} \right)
                  \biggr]
\end{equation}
with $0<c<\sqrt{n}$ and $c \neq \sqrt{n/2}$ and
\begin{equation}
       W^{(1)}_{NR} = \lambda _4 \left(
                      {N_1{\overline N} \over M_C^2}
                       \right) \left( {N_2{\overline N}
                                  \over M_C^2} \right)
                      +\lambda _5 \left( {N_3{\overline N}
                                  \over M_C^2} \right) ^2 \ .
\label{eqn:anr}
\end{equation}
The superfields $N_i(i=1,2,3)$ are contained in
$W^{(1)}_{NR}$.
The addition of $W^{(1)}_{NR}$ to $W^{(0)}_{NR}$
does not change the absolute minimum with the VEVs
$\langle N \rangle = \langle \overline N \rangle =
M_Cy_0$ and
$\langle S \rangle = \langle \overline S \rangle =
M_Cx_0$.
Here it is assumed that there is no term like
\begin{equation}
        \left( {S{\overline S} \over M_C^2} \right) ^{n/2}
                \left( {N_i{\overline N} \over M_C^2}
                       \right) \qquad (i=1,2,3).
\label{eqn:forbid}
\end{equation}
If $W_{NR}$ contains this type of the NR terms,
the VEVs at the absolute minima could change.
Absence of these terms can be guaranteed by
the introduction of discrete symmetries.
For illustration let us take here $n = 6$.
If the model contains the discrete symmetry
${\bf Z}_7 \times {\bf Z}_2$ and if each superfield has
a suitable discrete charge as shown in Table II,
the superpotential (\ref{eqn:NR}) to (\ref{eqn:anr}) is allowed
whereas the terms (\ref{eqn:forbid}) are forbidden.
In Table II the discrete charge of Grassmann number
is taken as $(-1,\  -)$.
It is interesting for us to remember Gepner model
in which Calabi-Yau manifold is constructed algebraically by
a tensor product of $N=2$ minimal superconformal models
with the level $k$'s
\cite{GEPNER}.
In Gepner model there appears the discrete symmetry
${\bf Z}_{k+2}\times {\bf Z}_2({\bf Z}_{k+2})$
for each $N=2$ minimal superconformal model
with an odd(even) level $k$.
In view of the fact that algebraic construction of
compactified manifolds brings about various types of
the discrete symmetry,
the present model is a likely scenario.

\vskip 0.5cm

\begin{center}
 \unitlength=0.8cm
 \begin{picture}(2.5,2.5)
  \thicklines
  \put(0,0){\framebox(3,1){\bf Table II}}
 \end{picture}
\end{center}

\vskip 0.5cm

{}From Eq. (\ref{eqn:anr}), we finally obtain the mass matrix
for the Majorana neutrino sector as
\begin{equation}
        M_M = \bordermatrix{
              & N_1 & N_2 & N_3 & N' & S' \cr
              & 0 & \sim \lambda_4M_{N'} & 0 & 0 & 0 \cr
              & \sim \lambda_4M_{N'} & 0 & 0 & 0 & 0 \cr
              & 0 & 0 & \sim \lambda_5M_{N'} & 0 & 0 \cr
              & 0 & 0 & 0 & M_{N'} & 0               \cr
              & 0 & 0 & 0 & 0 & \sim m_{SUSY}}  \ .
\end{equation}
So all Majorana neutrinos have masses of order $M_{N'}$
except for the field $S'$ which has the mass of
$O(m_{SUSY}) \simeq O(1 \tev)$.
Dirac-mass terms come from usual Yukawa interactions
\begin{equation}
      \lambda_{ij} N_i L_j H_{\rm u} \simeq
                    \lambda'_{ij} E_i L_j H_{\rm d}\,,
\end{equation}
where $E_i$ means $i$-th $SU(2)_L$-singlet charged
lepton fields.
Since $\langle H_{\rm u}\rangle \simeq \langle H_{\rm d}
\rangle \simeq O(10^2 \gev)$,
we obtain almost the same Dirac mass matrix
for neutrinos as for charged leptons.
So large M-masses induced by the above mechanism
yield very small neutrino masses via the see-saw mechanism.
The results are consistent with recent solar
and atmospheric neutrino experiments.


\section{Summary and Discussion}

In Calabi-Yau superstring models with abelian
flux breaking the gauge group is rank-six at
the compactification scale.
To connect Calabi-Yau models with the standard
model, there should exist two intermediate energy
scales of symmetry breaking between the compactification
scale and the electroweak scale.
In this paper, we clarified that
in a certain type of Calabi-Yau superstring models
the symmetry breaking occurs by stages
at two large intermediate energy scales
which are given by
$\langle S \rangle (\langle \overline S \rangle)$ and
$\langle N \rangle (\langle \overline N \rangle)$.
Two large intermediate scales induce a large M-masss of
right-handed neutrinos.
Peculiar structure of the effective NR interactions
is crucial in models.
Furthermore, the special sets
$m=2, l=1, n=2k \geq 6$ for the NR interactions
are necessary for viable scenarios,
in which the NR terms of the superpotential
become Eq. (\ref{eqn:WNR}).
In fact, the M-mass becomes $O(10^{9 \sim 10}\gev)$
for these cases and see-saw mechanism can be at work.
We proposed a concrete model with three generations
which leads to small see-saw M-masses for neutrinos.
This large M-mass solves the solar neutrino promlem and also
is compatible with the cosmological bound
for stable light neutrinos.
Special form of the NR terms suggests
that the superstring model possesses an appropriate
discrete symmetry coming from
distinctive structure of the compactified manifold.

Mass hierarchy of quarks and leptons may also have its origin
in the discrete symmetry and the presence of
large intermediate scales.
Provided that mirror superfields except for $S$ and $N$
are not contained in the model,
we may have distinct types of the NR terms, for instance,
associated with the up-quark sector as
\begin{equation}
      \sum _p \lambda ^{(p)}_{ij} \left(
                \frac {S \overline S}{M_C{}^2} \right)^p
                   Q_i U_j^c H_u\,,
\end{equation}
where $\lambda ^{(p)}_{ij} = O(1)$ and $Q_i$ and $U_j^c$
stand for quark-doublet and up-quark-singlet superfields
for the $i$-th generation, respectively.
If the discrete symmetry compel us to retain only
the terms
\begin{equation}
      \lambda ^{(2)}_{11}
               \left( \frac {S \overline S}{M_C{}^2} \right)^2
                   Q_1 U_1^c H_u +
      \lambda ^{(1)}_{22}
               \left( \frac {S \overline S}{M_C{}^2} \right)
                   Q_2 U_2^c H_u +
      \lambda ^{(0)}_{33} Q_3 U_3^c H_u
\end{equation}
for the up-quark sector of the superpotential,
we have the mass hierarchy of up-quarks such as
\begin{equation}
     m_u \sim \langle H_u \rangle x_0{}^4, \qquad
     m_c \sim \langle H_u \rangle x_0{}^2, \qquad
     m_t \sim \langle H_u \rangle.
\end{equation}
Since $x_0{}^2 \sim 10^{-(2.0 \sim 2.3)}$ for $n = 8$,
we obtain a plausible solution which is in accord with
experimental data.
At all events the effective NR interactions play
an important role in connecting the superstring theory
with the standard model.
It is the discrete symmetry of the compactified manifold
that controls the characteristic features of
the NR terms.



\newpage

{\large {\bf Table Captions}}

\vskip 2cm

{\bf Table I} \\
The energy scales of symmetry breaking $\langle S \rangle$
and $\langle N \rangle$ and a large Majorana-mass $M_{N'}$ in
GeV unit for various cases of $(n,k,m,l)$.
Here we take $M_C=10^{18.5}$GeV and $\sqrt {-m_S{}^2}=10^3$GeV.

\vskip 2cm

{\bf Table II} \\
The charge assignment of the discrete symmetry ${\bf Z}_7
\times {\bf Z}_2$ for superfields $S,\ \overline S,\ N,\
\overline N$, and $N_i$.
The discrete charge of Grassmann number is taken as
$(-1,\ -)$.
${\bf Z}_7 \times {\bf Z}_2$ is
taken as only an example of the discrete group.


\newpage

{\large {\bf Figure Captions}}
\vskip 2cm

{\bf Fig. 1} \\
The structure of the scalarpotential in
the case $(n,k,m,l)=(6,3,2,1)$
with $a=b=c=1$.
The vertical axis is taken as the normalized scalarpotential
$v$ (see text).
The horizontal axes are $\overline x = (x/x_0)^3$ and
$\overline y = y/y_0$, where $x=\langle S \rangle /M_C$
and $y= \langle N \rangle /M_C$. \\
(a) The overview of the scalarpotential $v$.
The Point A (the absolute minimum) is located at
$(\overline x, \overline y) = (1,1)$
and the Point B is a local minimum. \\
(b) The comparison of values of the scalarpotential $v$
between Point A and Point B.
A solid (dashed) curve represents
the calculation of $v$ vs. $\overline x$
along the line $\overline x=\overline y \ (\overline y=0)$. \\
(c) The comparison of $v$ vs. $\overline y$ along the line
with fixed $\overline x$-values.

\vskip 2cm

{\bf Fig. 2} \\
The structure of the scalarpotential in
the case $(n,k,m,l)=(9,3,3,2)$
with $a=b=1$ and $c=2$.
The vertical and horizontal axes are taken as the
same as in Fig.1. \\
(a) The overview of the scalarpotential $v$.
The Point A (the absolute minimum) is located at
$(\overline x, \overline y) = (1,1)$
and the Point B is a local minimum. \\
(b) The comparison of values of the scalarpotential $v$
between Point A and Point B.
A solid (dashed) curve represents
the calculation of $v$ vs. $\overline x$
along the line $\overline x=\overline y \ (\overline y=0)$. \\
(c) The comparison of $v$ vs. $\overline y$ along the line
with fixed $\overline x$-values.


\newpage


\begin{center}
{\bf Table I} \\
\bigskip

\begin{tabular}{|cccc|cc|c|}\hline
\quad  $n$   &   $k$   &   $m$   &   $l$ \quad
                & $\langle S \rangle$ (GeV)
                      &   $\langle N \rangle$ (GeV)
                            &   $M_{N'}$ (GeV)         \\ \hline
\hline
\quad
 4  &  2  &  2  &  1  \quad &  $10^{15.9}$  &  $10^{13.1}$
                                               &  $10^{8.1}$ \\
\quad
 6  &  3  &  2  &  1  \quad &  $10^{16.9}$  &  $10^{13.5}$
                                               &  $10^{8.8}$ \\
\quad
 8  &  4  &  2  &  1  \quad &  $10^{17.4}$  &  $10^{13.6}$
                                               &  $10^{9.1}$ \\
\quad
10  &  5  &  2  &  1  \quad &  $10^{17.6}$  &  $10^{13.7}$
                                               &  $10^{9.2}$ \\
\quad
12  &  6  &  2  &  1  \quad &  $10^{17.8}$  &  $10^{13.7}$
                                               &  $10^{9.2}$ \\
\quad
20  & 10  &  2  &  1  \quad &  $10^{18.1}$  &  $10^{13.7}$
                                            &  $10^{9.2}$ \\ \hline

\quad
 6  &  4  &  3  &  1  \quad &  $10^{16.7}$  &  $10^{14.7}$
                                               &  $10^{6.6}$ \\
\quad
 9  &  6  &  3  &  1  \quad &  $10^{17.5}$  &  $10^{15.3}$
                                               &  $10^{6.4}$ \\
\quad
12  &  8  &  3  &  1  \quad &  $10^{17.8}$  &  $10^{15.4}$
                                            &  $10^{6.6}$ \\ \hline

\quad
 6  &  2  &  3  &  2  \quad &  $10^{16.9}$  &  $10^{15.2}$
                                               &  $10^{5.4}$ \\
\quad
 9  &  3  &  3  &  2  \quad &  $10^{17.5}$  &  $10^{15.2}$
                                               &  $10^{5.8}$ \\
\quad
12  &  4  &  3  &  2  \quad &  $10^{17.8}$  &  $10^{15.3}$
                                            &  $10^{5.9}$ \\ \hline
\end{tabular}
\end{center}

\vspace{2cm}


\begin{center}
{\bf Table II} \\
\bigskip

\begin{tabular}{ | c | c | c |} \hline

\quad  Fields  \quad  &  ${\bf Z}_7$-charges  &
                        ${\bf Z}_2$-charges  \\ \hline

$S$             &     1       &     +        \\

${\overline S}$ &     1       &     +        \\ \hline

$N$             &     3       &     +        \\

${\overline N}$ &     3       &     +        \\ \hline

$N_1$           &     2       &    $-$       \\

$N_2$           &     4       &    $-$       \\

$N_3$           &     3       &    $-$       \\

$(\theta)$      &    $-1$     &    $-$       \\ \hline
\end{tabular}
\end{center}


\end{document}